\documentclass[sigconf]{acmart}
\renewcommand\footnotetextcopyrightpermission[1]{}
\usepackage{graphicx}
\usepackage{subfigure}
\usepackage{enumitem}
\usepackage{amsmath}
\usepackage{multirow}
\usepackage{tabularx}
\usepackage{soul}
\usepackage{array}
\usepackage{color}
\usepackage{adjustbox}
\newcolumntype{P}[1]{>{\centering\arraybackslash}p{#1}}
\makeatletter
\def\acm@copyrightmode{\relax}
\makeatother
 \usepackage{fancyvrb} 

\begin{document}
\title{Under the Shadow of Sunshine: Characterizing Spam Campaigns Abusing Phone Numbers Across Online Social Networks}

\author{Srishti Gupta}
\affiliation{%
  \institution{IIIT-Delhi}
  }

\author{Dhruv Kuchhal}
\affiliation{%
  \institution{MAIT, Delhi}
  }

\author{Payas Gupta}
\affiliation{%
  \institution{Pindrop, Atlanta}
  }

\author{Mustaque Ahamad}
\affiliation{%
  \institution{Georgia Institute of Technology}
  }

\author{Manish Gupta}
\affiliation{%
  \institution{Microsoft, India}
  }

\author{Ponnurangam Kumaraguru}
\affiliation{%
  \institution{IIIT-Delhi}
  }

\begin{abstract}
Cybercriminals abuse Online Social Networks (OSNs) to lure victims into a variety of spam. Among different spam types, a less explored area is OSN abuse that leverages the telephony channel to defraud users. Phone numbers are advertized via OSNs, and users are tricked into calling these numbers. To expand the reach of such scam~/~spam campaigns, phone numbers are advertised across multiple platforms like Facebook, Twitter, GooglePlus, Flickr, and YouTube.
In this paper, we present the first data-driven characterization of cross-platform campaigns that use multiple OSN platforms to reach their victims and use phone numbers for monetization. 

We collect $\sim$23M posts containing $\sim$1.8M unique phone numbers from Twitter, Facebook, GooglePlus, Youtube, and Flickr over a period of six months. Clustering these posts helps us identify 202 campaigns operating across the globe with Indonesia, United States, India, and United Arab Emirates being the most prominent originators. We find that even though Indonesian campaigns generate highest volume~($\sim$3.2M posts), only 1.6\% of the accounts propagating Indonesian campaigns have been suspended so far. 
By examining campaigns running across multiple OSNs, we discover that Twitter detects and suspends $\sim$93\% more accounts than Facebook. Therefore, sharing intelligence about abuse-related user accounts across OSNs can aid in spam detection. According to our dataset, around $\sim$35K victims and $\sim$\$8.8M could have been saved if intelligence was shared across the OSNs.
By analyzing phone number based spam campaigns running on OSNs, we highlight the unexplored variety of phone-based attacks surfacing on OSNs.
\end{abstract}

\maketitle
\section{Introduction}
The increasing popularity of Online Social Networks (OSNs) has attracted a cadre of criminals who craft large-scale phishing and spam campaigns targeted against OSN users. 
Traditionally, spammers have been driving traffic to their websites by luring users to click on URLs in their posts on OSNs~\cite{grier2010spam, gao2010detecting, thomas2011suspended}. 
A significant fraction of OSN spam research has looked at solutions driven by URL blacklists~\cite{gao2010detecting, thomas2011design}, manual classification~\cite{benevenuto2009detecting}, and honeypots~\cite{lee2010uncovering, stringhini2010detecting}. 
Since defence mechanisms against malicious~/~spam URLs have already matured, 
cybercriminals are looking for other ways to engage with users. Telephony has become a cost-effective medium for such engagement, and phone numbers are now being used to drive call traffic to spammer operated resources (e.g., call centers, Over-The-Top messaging applications like WhatsApp).

In this paper, we explore a data-driven approach to understand OSN abuse that makes use of phone numbers as action tokens in the realization~/~monetization phase of spam campaigns. Telephony has turned out to be an effective tool for spammers because Internet crime reports suggest that people fell victim to phone scams leading to a loss of \$7.4B in 2015 for Americans alone~\footnote{\url{https://blog.truecaller.com/2017/04/19/truecaller-us-spam-report-2017/}}. 
Specifically, in the phone-based abuse of OSNs, spammers advertise phone numbers under their control via OSN posts and lure OSN users into calling these numbers. Since spammers use phone calls to trap victims, it is safe to assume that spammers would provide real phone numbers under their control. 
In addition, advertising phone numbers reduce spammers' overhead of finding the set of potential victims who can be targeted via the phone. Over phone conversations, they try convincing the victims that their services are genuine, and deceive them into making payments~\cite{miramirkhani2017dial}. To maximize their reach and impact, we observe that spammers disseminate similar content across multiple OSNs. 

While URLs help spammers attract victims to websites that host malicious content, phone numbers provide more leverage to spammers. Due to the inherent trust associated with the telephony medium and the impact of human touch over phone calls, spammers using phone numbers stand a better chance of convincing and hence are likely to make more impact. Besides, they can use fewer phone numbers as compared to URLs; a large number of URLs are required to evade filtering mechanisms incorporated by OSNs.~\footnote{\url{https://support.twitter.com/articles/90491}} 
Moreover, the monetization and advertising channel in phone-based campaigns i.e.,~(Phone) and~(Web) respectively is different as compared to a single channel~(Web) used in URL-based campaigns. Hence, phone-based spam requires correlation of abuse information across channels which makes it harder for OSN service providers to build effective solutions. 
Since the modus operandi in URL-based and phone-based spam campaigns is different, leaving phone-based spams unexplored can limit OSN service providers' ability to defend their users from spam. 
While extensive solutions have been built to educate users about URL-based spam~\cite{kumaraguru:anti-phishing-landing-page:2009:yuqfj}, limited education is available for phone-based attacks.
This is evident from several well publicized and long running Tech Support spam campaigns (since 2008) that use phone numbers to lure victims leading to huge financial losses in the past, as reported by the Federal Bureau of Investigation~\cite{techsupportfbi}. Although detecting and avoiding OSN abuse using phone numbers is so critical now, to the best of our knowledge, this space is largely unexplored. 

In this paper, we address this gap by taking the \emph{first} step in \emph{identifying} and \emph{characterizing} spam campaigns that abuse phone numbers across multiple OSNs. Studying phone-based spam across multiple OSNs provides a new perspective and helps in understanding how spammers work in coordination to increase their impact. From 22M posts collected from Twitter, Facebook, GooglePlus, YouTube, and Flickr, we identify 202 campaigns running across different countries, leveraging 806 unique abusive phone numbers. 
Studying these campaigns, we make the following key observations:
\begin{enumerate}
\item We find that the cross-platform phone based spam campaigns originate from more than 16 countries, but most of them come from Indonesia, United States of America~(USA), India, and United Arab Emirates (UAE). These campaigns are supported less number of phone numbers as compared to URLs, perhaps due to (a) the high cost of acquiring a phone number, and (b) weak defense mechanisms against phone~-~based spam. Victims that fall prey to these campaigns are offered banned filmography, personal products and a variety of other services; but the services are not delivered even after successful payment.
\item As reported in earlier research~\cite{ghosh2012understanding}, we also find evidence that suggests spammers collude to maximize their reach either by creating multiple accounts or promoting other spammers' content. To evade suspension strategies of each OSN, spammers keep the volume per account low. Our results show that accounts are suspended after being active for 33 days (on average); while literature suggests that spammers involved in URL-based spam campaigns, on the other hand, could survive only for three days after their first post~\cite{thomas2011suspended}. In addition, 68.7\% of spammer accounts are never suspended. Again, this suggests a crucial need to build effective solutions to combat phone-based spam.
\item Our analysis also suggests that OSN service providers should work together in the fight against phone-based spam campaigns. By examining phone numbers involved in campaigns across OSNs, we find that although all OSNs are consistently being abused, Twitter is the most preferred OSN for propagating a phone campaign. 
By analyzing spammers' multiple identities across OSNs, we find that Twitter is able to suspend 93.3\% more accounts than Facebook. Thus, \emph{cross-platform intelligence} can be useful in preventing the onset and reducing the lifetime of a campaign on a particular network with good accuracy. We estimate that cross-platform intelligence can help protect 35,407 victims across OSNs, resulting in potential savings of \$8.8M.
\end{enumerate}

Altogether, our results shed light on phone-based spam campaigns where spammers are using one channel~(OSN) to spread their content, and the other channel~(voice~/~SMS~/~message via phone) to convince their victims to fall prey to their campaigns. Given that no timely and effective filters exist on either channel to combat such spam, there is an imperative need to build one. 
\section{Related Work}
Spam is a growing problem for OSNs, and several researchers have looked at different ways to combat it. In this section, we present prior research in detecting spam campaigns on OSNs.

\textbf{Handling non-phone based spam:} 
There has been a large body of work that reports the existence of spam on multiple OSNs like YouTube~\cite{benevenuto2009detecting}, Twitter~\cite{grier2010spam}, and Facebook~\cite{gao2010detecting}. 
Thomas et al. studied the characteristics of suspended accounts on Twitter~\cite{thomas2011suspended}. With an in-depth analysis of several spam campaigns, they reported that 77\% spam accounts suspended by Twitter were taken down on the day of their first tweet. Apart from this, there has been work done to differentiate a spammer from a non-spammer~\cite{yardi2009detecting, benevenuto2010detectingtw, wang2010don, lee2011seven, amleshwaram2013cats}.
Lumezanu et al. studied the spread of URL campaigns on email and Twitter and found that spam domains receive better coverage when they appear both on Twitter and email~\cite{lumezanu2012observing}. 
In addition to characterizing URL-based spam, methods have been proposed for detecting~\cite{lee2010uncovering, webb2008social, chu2012detecting} and preventing~\cite{rahman2012frappe, faloutsos2013detecting} such campaigns.
While a lot of work has been done on characterizing and detecting URL-based spam campaigns, campaigns abusing phone numbers have been largely ignored. 

\textbf{Handling phone based spam:} A large fraction of phone spam includes robocalling and spoofing, wherein spammers call the victims and trick them into giving personal or financial information~\footnote{\url{https://www.consumer.ftc.gov/articles/0076-phone-scams}}.
Studies have shown that, in spam activities, phone numbers are more stable over time than email, and hence can be more helpful in identifying spammers~\cite{costin2013role, isacenkova2014inside}. Christin et al. analyzed a type of scam targeting Japanese users, threatening to reveal the users' browsing history, in case they do not give them money~\cite{christin2010dissecting}.
In studies mentioned above, the authors relied on publicly available datasets to perform their analyses. In contrast, we develop an infrastructure to collect millions of posts from OSNs, cluster them into campaigns, and conduct our analyses.
Researchers have investigated phone number abuse by analyzing cross-application features in Over-The-Top applications~\cite{gupta2016exploiting}, cross-channel SMS abuse~\cite{srinivasan2016understanding}, and by characterizing honeypot numbers~\cite{gupta2015phoneypot,payas_m3aawg,mobipot2016, marzuoli2016uncovering}. 
Recently, Miramirkhani et al. studied the Tech Support campaign that abuse phone numbers, from the perspective of domains that were used to host malicious content~\cite{miramirkhani2017dial}. The authors also interacted with spammers to understand their social engineering tactics. While they focused on URLs and domains abused by spammers, we study the cross-platform spread of phone-based spam campaigns across OSNs, along with strategies adopted by spammers for sustainability and visibility. Besides, we highlight how cross-platform intelligence about spam accounts can be shared across OSNs to aid in spam detection.
\section{Dataset} \label{dataset}

\begin{figure}[h]
\begin{center}
\includegraphics[width=0.95\linewidth]{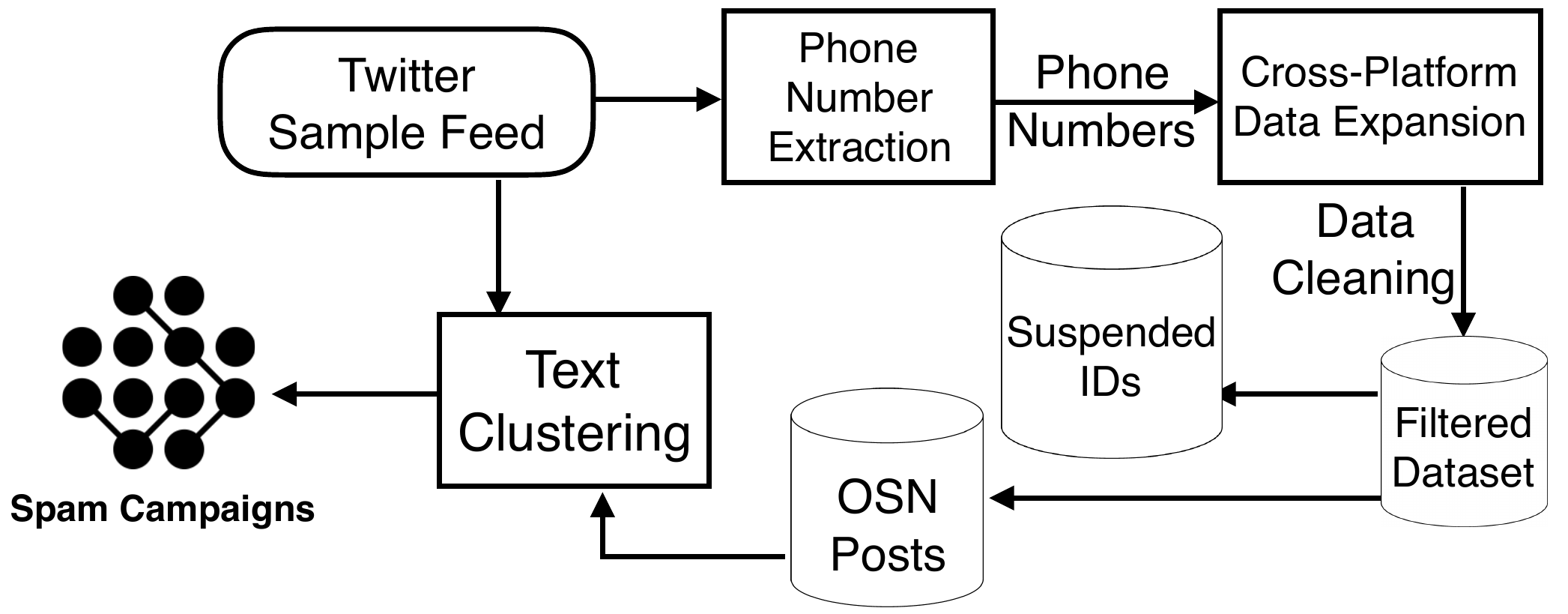}
\caption{System Architecture for Data Collection across Multiple OSNs.}
\label{fig:osn-vol}
\end{center}
\end{figure}

In this section, we discuss our methodology for collecting phone numbers, posts and other metadata; which we use later to find campaigns on OSNs. These campaigns are then tagged as benign or spam.
Figure~\ref{fig:osn-vol} shows the architecture of our data collection subsystem that is used to collect phone numbers across multiple OSNs. We picked Twitter as the starting point to find phone numbers, as it provides easier access to large amounts of data as compared to other online social networks~\cite{osborne2014facebook}. 
We set up a framework to collect a stream of tweets containing phone numbers. Some of the keywords used in data collection and regular expressions used to extract phone number from a text are listed in the Appendix~\ref{regex}. For each unique phone number received every day, a query was made to other OSNs viz. Facebook,~\footnote{Collecting data from Facebook was challenging. In April 2015, Facebook deprecated their post-search API end-point~\footnote{\url{https://developers.facebook.com/docs/graph-api/using-graph-api/v2.0}}, so we used an Android mobile OAuth token to search content using the Graph API~\cite{gupta2016exploiting}.} GooglePlus, Flickr, and YouTube, and for every search, we stored the following details: user details (user ID, screen name, number of followers and friends), post details (time of publication, text, URL, number of retweets, likes, shares, and reactions), and whether the ID were suspended. The data collection ran over a period of six months, between April 25, 2016 and October 26, 2016. Our system collected 22,690,601 posts containing 1,845,150 unique phone numbers, posted by 3,365,017 unique user accounts on five different OSNs. After removing noise~(i.e., the posts which do not contain a phone number), the filtered set was used for finding campaigns.

We acknowledge that our dataset may contain two kinds of bias: (1) Only 1\% sample of all public tweets is available from the Twitter Streaming API; it can underestimate the spam campaigns observed on Twitter. (2) Since we treat Twitter as the starting point, we may miss some campaigns which are popular on other social networks, but not on Twitter. However, Twitter provides best access to user posts, justifying our choice.  

\textbf{Campaigns:} 
A \emph{campaign} is defined as a collection of posts made by a set of users sharing similar text and phone numbers. 
To make sure that we do not tag any benign campaign as spam, we filtered out the phone numbers used by even one Twitter verified account. 
Every phone number, say \emph{ph1}, is represented by a set of frequent unigram tokens which occur around the phone number. All posts that contain at-least 33\% tokens from the representative token set are put together in a cluster; indicating posts related to the phone number. Different phone numbers, say \emph{ph1} and \emph{ph2}, are put together in the same cluster if the average Jaccard coefficient between the corresponding set of posts is greater than 0.7. We calculated different values of Jaccard coefficient and average silhouette scores to measure quality of clusters~\cite{almeida2011there}, and found 0.7 as knee point for corresponding value of silhouette score as 0.8. All users that post about any phone number in the clustered set are put together. A cluster thus formed is marked as a campaign. Using this method, we found 22,390  campaigns in the dataset, collectively amounting to $\sim$10.9M posts.

\textbf{Spam Campaigns:} We flag a campaign as \emph{spam} if it meets the following criteria: (a) phone number involved in the campaign is present in the United States Federal Trade Commission's Do Not Call (DNC) dataset~\footnote{\url{https://www.ftc.gov/site-information/open-government/data-sets/do-not-call-data}}, or (b) even if one OSN account involved
in the campaign is suspended. Further, to be able to characterize the spam campaigns in detail, we focused only on campaigns with at least 5000 posts. With this, we identified 6,171 out of 22,390 campaigns as spam. From this set of campaigns, we did a manual inspection to verify if the campaign is indeed spam. 
 This results in a working dataset of \emph{202 campaigns} comprising of \emph{$\sim$4.9M posts}. During manual inspection, we also  assigned topics to the 202 campaigns, where multiple campaigns could be assigned the same topic. 
For instance, a campaign selling shoes and other selling jackets would be assigned the topic -- ``Product Marketing". 
\section{Characterizing Spam Campaigns} \label{allcampaigns}
In this section, we focus on the following research questions. Where do spam campaigns originate from? Do spammers use automation when posting phone numbers or answering ``phone calls''? What does a spammer OSN account suspension depend on? What is the typical modus operandi of the spammers?

\subsection{Where does Phone-based Spam Originate?}
It is important to know from which countries does the spam originate; it can be used in developing anti-spam filtering solution. We assume that the country associated with a phone number is the source country. For the analysis, we need to extract the country of the spam phone number. This is done either by identifying (a) the language of the post containing the spam phone number via the `lang' field in the tweet object, or (b) by the country code using Google's phone number library.~\footnote{https://github.com/googlei18n/libphonenumber} These two methods helped in identifying countries for 127 campaigns. For rest of the campaigns, we called up the top two frequently occurring phone numbers in the campaign using Tropo~\footnote{https://www.tropo.com/}, a VoIP software that can be used to make spoofed calls. We recorded all the calls and used Google's Speech API~\footnote{https://cloud.google.com/speech/} to detect language and country of the campaign. We could identify origin country for 26 more campaigns; for the remaining 49, the country is unknown. Table~\ref{all_campaigns} presents topic distribution across various campaigns originating from different countries along with the average number of posts being made in each campaign. While majority of the spam was similar to advanced-fee scam~\footnote{}, where spammers trick victims to make payments in advance, there were certain different type of campaigns observed in the dataset as well: Hacking~(Tech Support) and Alternating Beliefs~(Love Guru). In the \emph{LoveGuru} campaign, astrologers promise victims to fix their love and marriage related problems. 
In the \emph{Tech Support} campaign, spammers pose as technical support representatives or claim to be associated with big technological companies (like Amazon, Google, Microsoft, Quebec, Norton, Yahoo, Mcafee, Dell, HP, Apple, Adobe, TrendMicro, and Comcast) and offer technical support fixes.

\begin{table}[!h]
\small{
\centering
\caption{Distribution of Campaigns across Topics and Source Countries. (\#C denotes the number of campaigns).}
\label{all_campaigns}
\begin{tabular}{|l|p{1.8in}|c|c|}
\hline
Country   & Campaign Topics& \#C& \#Posts                                                                                                                 \\ \hline
Argentina & \begin{tabular}[c]{@{}l@{}}Party Reservations\\ Pornography\end{tabular}                                                                                                                                                                                 & \begin{tabular}[c]{@{}c@{}}1\\ 1\end{tabular}                                   & \begin{tabular}[c]{@{}c@{}}39,476\\ 30,751\end{tabular}                                                                     \\ \hline
Chile     & Delivering Goods                                                                                                                                                                                                                                          & 1                                                                               & 6,691                                                                                                                       \\ \hline
Columbia  & \begin{tabular}[c]{@{}l@{}}Hotel Booking\\ Pornography\end{tabular}                                                                                                                                                                                      & \begin{tabular}[c]{@{}c@{}}1\\ 1\end{tabular}                                   & \begin{tabular}[c]{@{}c@{}}18,228\\ 5,324\end{tabular}                                                                      \\ \hline
Ghana     & \begin{tabular}[c]{@{}l@{}}Alternating Beliefs~(Marriage, Anxiety)\end{tabular}                                                                                                                                                                         & \begin{tabular}[c]{@{}c@{}} 2\end{tabular}                                   & 12,825                                                                                                                      \\ \hline
Guatemala & Product Marketing                                                                                                                                                                                                                                         & 1                                                                               & 8,821                                                                                                                       \\ \hline
India     & \begin{tabular}[c]{@{}l@{}}Hotel Booking\\ Alternating Beliefs~(Marriage, Anxiety)\\ Hacking(Tech Support)\end{tabular}                                                                                                                                           & \begin{tabular}[c]{@{}c@{}}1\\ 1\\ 1\end{tabular}                            & \begin{tabular}[c]{@{}c@{}}10,986\\ 15,128\\ 43,552\end{tabular}                                                         \\ \hline
Indonesia & \begin{tabular}[c]{@{}l@{}}Hotel Booking\\ Product Marketing\\ Pornography\\ Alternating Beliefs~(Marriage, Anxiety))\\ Purchasing Followers\\Finance, Real Estate\\ Selling Adult Products\\ Uncategorized\end{tabular}                                            & \begin{tabular}[c]{@{}c@{}}1\\ 75\\ 4\\ 7\\ 15\\ 3\\ 5\\ 3\end{tabular}      & \begin{tabular}[c]{@{}c@{}}8,291\\ 2,689,616\\ 164,382\\ 101,799\\ 406,713\\ 23,700\\ 48,109\\ 29,043\end{tabular}       \\ \hline
Kuwait    & Charity (Donation)                                                                                                                                                                                                                                        & 1                                                                               & 46,494                                                                                                                      \\ \hline
Mexico    & Pornography                                                                                                                                                                                                                                               & 1                                                                               & 8,204                                                                                                                       \\ \hline
Nigeria   & \begin{tabular}[c]{@{}l@{}}Alternating Beliefs~(Marriage, Anxiety)\end{tabular}                                                                                                                                                                         & 1                                                                               & 29,226                                                                                                                      \\ \hline
Pakistan  & Finance, Real Estate                                                                                                                                                                                                                                      & 1                                                                               & 16,058                                                                                                                      \\ \hline
Spain     & Charity (Donation)                                                                                                                                                                                                                                        & 1                                                                               & 14,311                                                                                                                      \\ \hline
UAE       & Escorts                                                                                                                                                                                                                                               & 5                                                                               & 69,263                                                                                                                      \\ \hline
USA       & \begin{tabular}[c]{@{}l@{}}Party Reservations\\ Product Marketing\\ Pornography\\ Alternating Beliefs~(Marriage, Anxiety)\\ Escorts\end{tabular}                                                                                                         & \begin{tabular}[c]{@{}c@{}}8\\ 1\\ 1\\ 1\\  1\end{tabular}                    & \begin{tabular}[c]{@{}c@{}}172,090\\ 22,804\\ 19,653\\ 12,936\\  9,652\end{tabular}                                       \\ \hline
UK        & \begin{tabular}[c]{@{}l@{}}Escorts\\ Charity (Donation)\end{tabular}                                                                                                                                                                                      & \begin{tabular}[c]{@{}c@{}}1\\ 2\end{tabular}                                   & \begin{tabular}[c]{@{}c@{}}9,268\\ 17,184\end{tabular}                                                                      \\ \hline
Venezuela & \begin{tabular}[c]{@{}l@{}}Hotel Booking\\ Free Games, Downloads\end{tabular}                                                                                                                                                                             & \begin{tabular}[c]{@{}c@{}}1\\ 1\end{tabular}                                   & \begin{tabular}[c]{@{}c@{}}6,813\\ 9,028\end{tabular}                                                                       \\ \hline
Unknown   & \begin{tabular}[c]{@{}l@{}}Party Reservations\\ Hotel Booking\\ Product Marketing\\ Free Games, Books, Downloads\\ Pornography\\  Alternating Beliefs~(Marriage, Anxiety)\\ Finance, Loans, Real Estate\\ Charity (donation)\\Uncategorized\end{tabular} & \begin{tabular}[c]{@{}c@{}}10\\ 2\\ 10\\ 1\\ 17\\ 5\\  2\\ 2\\ 2\end{tabular} & \begin{tabular}[c]{@{}c@{}}323,565\\ 11,334\\ 108,634\\ 8,834\\ 211,714\\ 48,093 \\ 34,226\\ 29,740\\ 10,266\end{tabular} \\ \hline
\end{tabular}}
\end{table}


Top four source countries selected by the volume of campaigns viz. Indonesia, United States of America (USA), India, and United Arab Emirates (UAE) show interesting characteristics.
From Table~\ref{all_campaigns}, we observe that there is a good overlap of campaign categories across countries, while some countries have specific categories of campaigns running. Among all the campaign categories, volume generated by Indonesian campaigns is significantly higher than any other country.

\subsection{Do Spammers use Automation?}
While investigating further, we found that 99.3\% pairs of consecutive posts related to the same campaign appeared on Twitter in less than 10 minutes. Given that a major fraction of content appeared within a few minutes,
it is likely that content generation is automated. To ascertain
this, we looked at the information of the client (provided by the Twitter API) used by spammers to interact with the Twitter API or their
web portal. We found that most of the content was generated using
`twittbot.net', a popular bot service, known to be used by spammers~\cite{thomas2011suspended}. Apart from the bot service, several other clients like RoundTeam (0.25\%), IFFTT (0.03\%), Buffer (0.017\%), and Botize (0.016\%), were used for Twitter. Besides, we found that volume per phone number was also high in Indonesian campaigns; 80\% phone numbers had more than 1000 posts. One would assume that volume per phone number would be low since there are humans at the other end to service the requests. However, by processing the text in the posts created in this campaign, we found that spammers requested users to communicate via SMS or WhatsApp~($\sim$ 71\% posts). This explains why spammers would be able to handle the load of interacting with victims. There are many other advantages of using these messaging services -- spammers can further send phishing messages to victims, communicate with them unmonitored, and potentially use automated bots to reply to SMSs or Whatsapp messages. 

\subsection{What Factors Govern Spammers' Suspension?}

As expected, we find that the visibility (number of likes, shares, and retweets) of a post is positively correlated with the number of posts (Pearson correlation coefficient = 0.97). While this may sound intuitive, the number of accounts that were suspended within a campaign were not positively correlated with the number of posts. We noticed that even though the volume generated by Indonesian campaigns was 98.2\% higher than Indian campaigns, the fraction of users suspended in Indian campaigns was 85.6\% higher. Further, we observed that the account suspension is dependent on the nature of campaigns; campaigns providing escort services or technical support services had more accounts suspended. 

Surprisingly, for similar escort service campaign running in two different countries, USA and UAE, there was a significant difference in the number of accounts suspended. Before concluding that the country plays a major role in account suspension, we performed detailed analysis as follows.

\begin{figure}[h]
\begin{center}
\includegraphics[width=8cm,height=8cm,keepaspectratio]{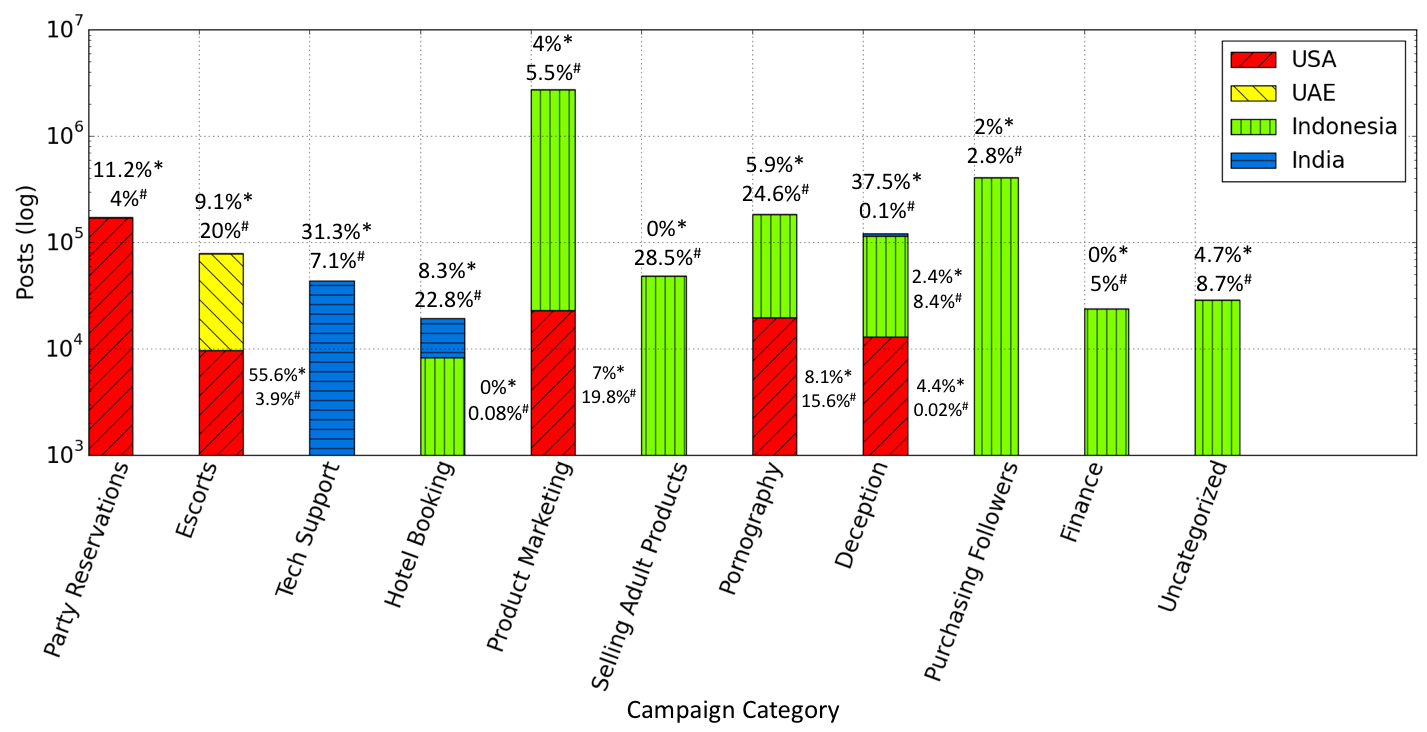}
\caption{Comparison of campaigns running in the top 4 countries -- Indonesia, USA, India, and UAE across different campaign categories. While visibility that a post receives is positively correlated with volume, account suspension in a campaign is not. Escort service and Tech Support campaigns had largest percentage of suspended accounts. The number of users suspended is represented by * and \# denotes the fraction of posts getting visibility.}
\label{fig:osnvol}
\end{center}
\end{figure}

\begin{figure}[h]
\subcapraggedrighttrue
\begin{center}
\subfigure[]
{ \label{fig:cmp_vol}
\includegraphics[width=4cm,height=4cm,keepaspectratio]{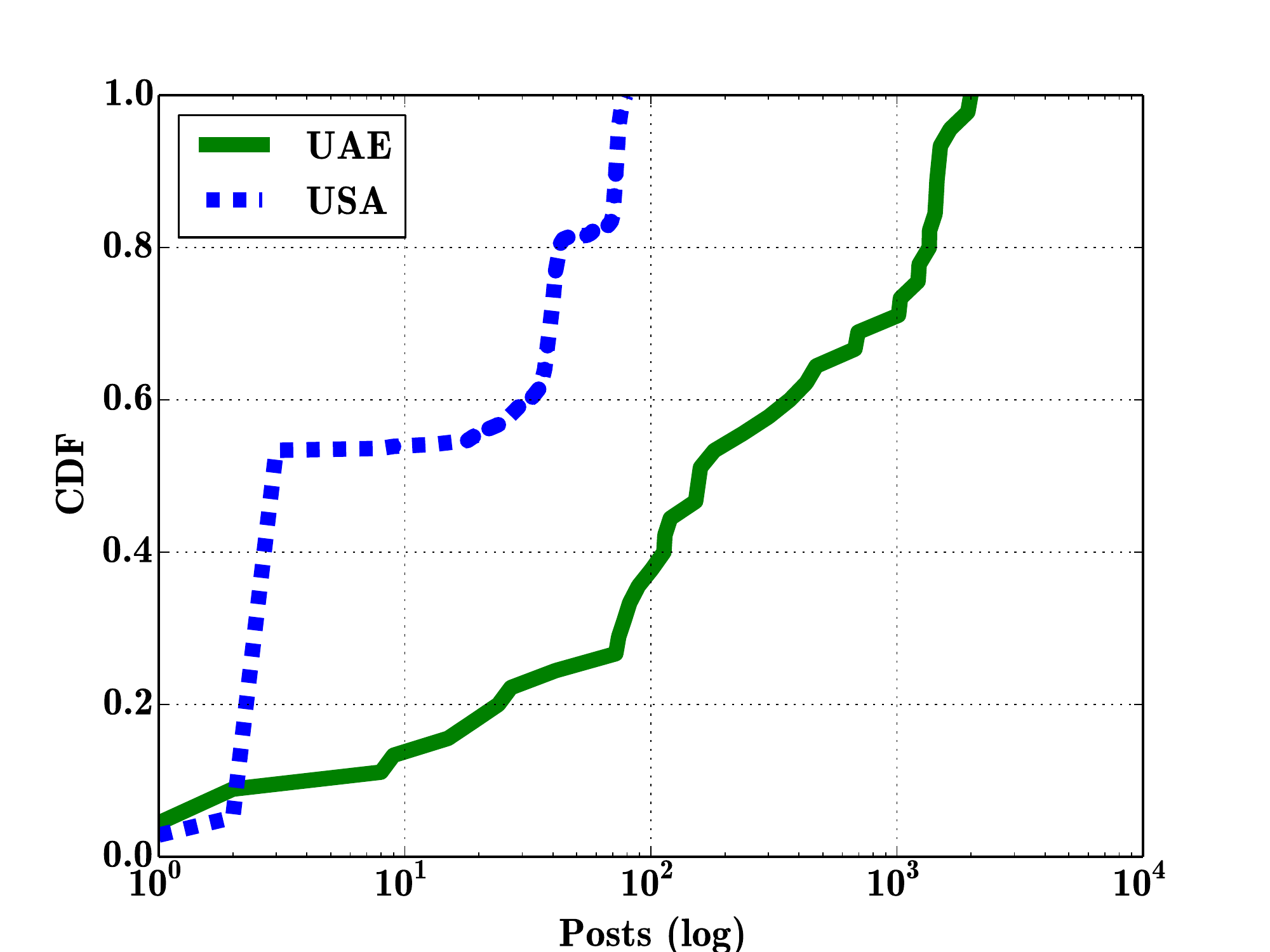}
}
\subfigure[]
{ \label{fig:cmp_interarrival}
\includegraphics[width=4cm,height=4cm,keepaspectratio]{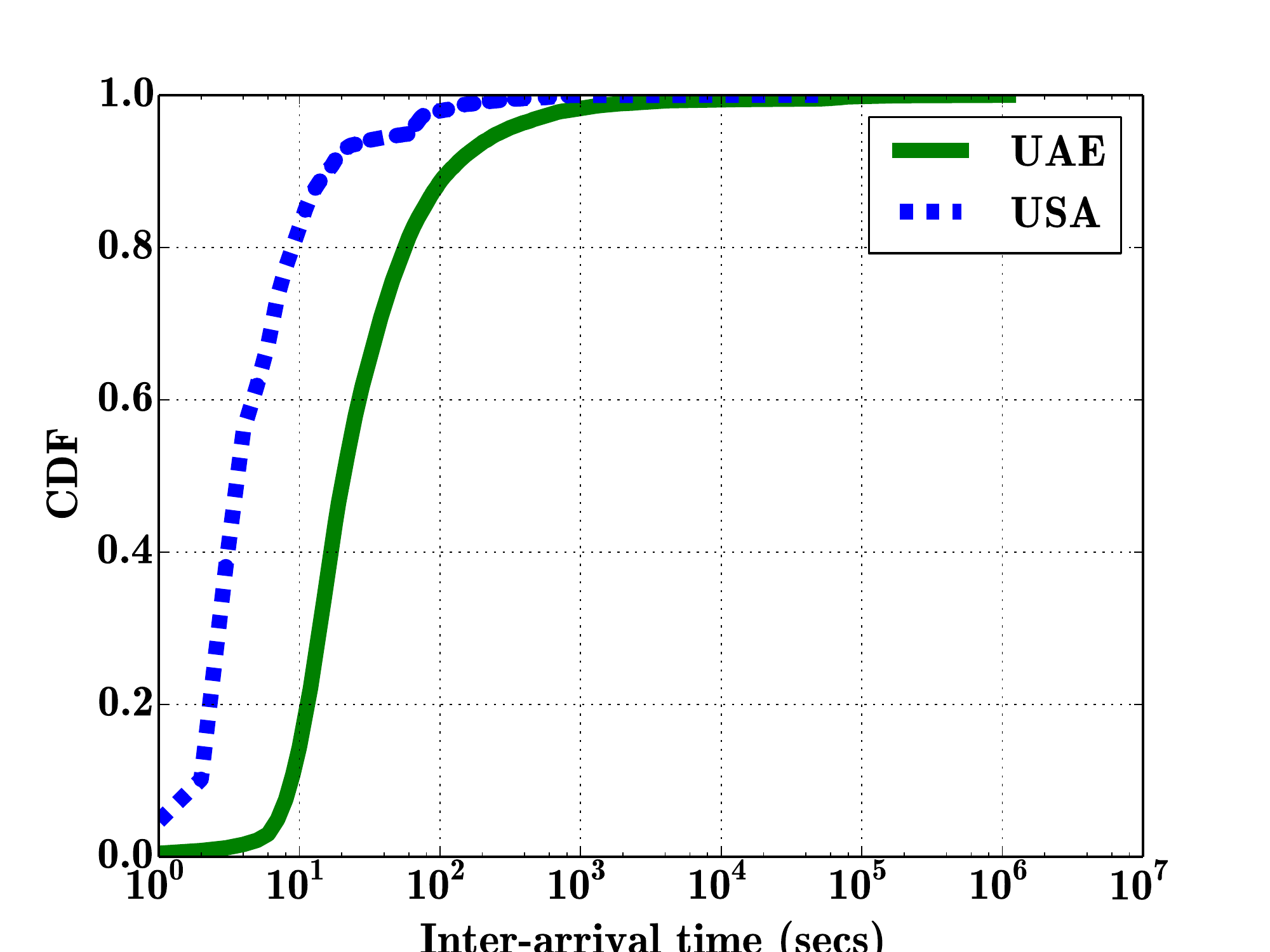}
}
\caption{Comparing Escort service campaign in USA vs. UAE. Even though volume generated per USA account is lower than UAE accounts~(a), inter-arrival time between two consecutive posts in the USA is lesser which could be a potential reason for suspension of accounts~(b).}
\label{fig:allcampaigns}
\end{center}
\end{figure}

The number of posts generated by escort campaign running in the USA (9,652) was lower than that running in UAE (69,263), but 55.6\% user accounts were suspended in the USA in comparison to only 9.1\% accounts suspended in UAE. We looked at several reasons which could potentially lead to account suspension -- volume generated per user or URLs used in the posts. We noticed that volume per user was higher for UAE users~(Figure~\ref{fig:cmp_vol}), number of URLs shared in UAE campaign was higher, and words used in both the campaigns had a good overlap. 
Also, from Figure~\ref{fig:cmp_interarrival}, we observed that inter-arrival time between two consecutive posts made by all the users in the USA~(41s on an average) is lesser than that of posts made in the UAE campaign~(392s on an average). 

\subsection{What is the Spammers' Modus Operandi?}
To ascertain the attack methodology the victims faced, we performed an experiment after receiving our institute's Institutional Review Board (IRB) approval. 
Pretending to be a potential victim, we called up phone numbers mentioned in campaigns selling adult (Viagra) pills in USA and UAE. In Indonesia, we interacted with spammers selling herbal products, and in India with those promoting tech support and astrology services~(providing solutions to marriage and love problems). To
avoid time zone conflict, we called the spammers in their local time of the day. Overall, we made 41 calls to different phone numbers from Indonesia, India, USA and UAE. A transcript of an interaction with a USA based spammer selling Viagra pills is listed in the Appendix~\ref{transcript}. Apart from Indonesia, campaigns from other countries had an IVR deployed, before reaching a spammer. We posit this can help in load balancing between limited human resources on the spammers' end. Due to language limitation in Indonesia, spammers preferred chatting over platforms like WhatsApp, where they were extremely responsive. 
 
The campaigns in USA and UAE were not limited by any delivery location; they had a usual delivery time of 2--4 weeks. These campaigns were operating solely over the phone and had no option of visiting an online portal to make the transaction. The attackers confidently asked for the credit card details over the phone even though banks advise otherwise. Spammers from Indonesia told that they would start delivery only after receiving the payment, which was to be done via bank transfer. During the interactions, spammers were persuasive in selling products by claiming their products to be the best as compared to similar products in the market. Tech support campaigns in India were providing service to users remotely over the Internet and charged over call once the issue was `fixed'. The catch was that the spammers pretended that there was a problem with the victims' computer and then tried to convince the victim to pay them to fix it, as reported in several complaints~\footnote{\url{https://800notes.com/Phone.aspx/1-800-549-5301/2}}. Another astrology based spam campaign running in India tricked by promising to fix users' marriage and love related problems within 48 hours~\footnote{\url{https://www.complaintboard.in/complaints-reviews/vashikaran-fake-vashikaran-fraud-cheater-money-taker-l149781.html}}. We called 4 numbers in different Indian states. Interestingly, all the spammers had a similar way of dealing with the problem, where they asked to send personal details over WhatsApp. 

It is evident that spammers running campaigns in different countries deploy similar mechanisms to let the victim reach them (posts on social media), to set up the product~/~service delivery operation (product delivery post payment and service delivery prior to payment), and model of payment (details transfer via phone, WhatsApp, verbal). It is the product delivery operation that creates deliberate confusion for a victim; intuitively, the delivery mechanism is similar for benign campaigns. Spammers leverage the advantage of similar delivery mechanisms, offer fake promises and later do not deliver.
\section{Characterizing Cross-Platform Spam Campaigns} \label{crossplatform}
In this section, we aim to answer the following research questions. Are spam campaigns run in a cross-OSN manner? How does the content cross-pollinate across OSNs? How do spammers maximize visibility? To what extent OSNs are able to detect phone based spam? Can existing intelligence on URL based spam be trivially adapted to handle the growing phone based spam problem? Can cross-platform intelligence help?

\subsection{Do Phone-based Spam Campaigns run in a Cross-OSN Manner?}
We observed that spam campaigns do not limit themselves to one OSN and are rather present on multiple networks. The distribution of posts across platforms in top 3 spam campaigns: Loveguru (from Alternating Beliefs category), Tech Support, and Indonesian Herbal Product (from Product Marketing category) is shown in Table~\ref{topcrossplatform}.
Even though Twitter has the largest fraction (possibly thanks to the first data source bias in our data collection method), all OSNs are abused to carry out spam campaigns. 
\begin{table}[h]
\centering
\caption{Top Cross-Platform Spam Campaigns}
\label{topcrossplatform}
\footnotesize
\setlength\tabcolsep{3pt}
\begin{tabular}{| P{3cm} | P{1.16cm} | P{0.66cm} | P{0.66cm} | P{0.65cm} | P{0.66cm} |} \hline
Campaign & TW & FB & G+ & YT & FL \\ \hline\hline
Tech Support & 28,984 & 2,151 & 7,830 & 2,850 & 1,737\\ \hline
LoveGuru & 6,934 & 1,418 & 4,257 & 101 & 63\\ \hline
Indonesia Herbal Product & 1,443,619 & 9,238 & 21 & 46 & 336\\\hline
\end{tabular}
\end{table}

Due to lack of space, in this section, we focus on studying in detail the Tech Support campaign. The details for other campaigns are available at~\textit{\url{http://bit.ly/phcamp-dash}}. Tech support scams have been around for a long period~\footnote{https://blog.malwarebytes.com/tech-support-scams/},incurring financial losses of \$2.2M to victims in 2016 alone, as reported by the US Federal Bureau of Investigation (FBI)~\cite{techsupportfbi}. Earlier, attackers used to call victims offering to fix their computer or PC. Now, attackers have changed their strategy; instead of calling victims, attackers float their phone numbers on OSNs and ask users to call them in case they need any technical assistance related to their computers. Once the victim calls the phone number, the
attacker asks for remote access to their machine to diagnose
the problem. The attacker fudges the expected problems with
victim`s machine and convinces her to get it fixed.
The reason this campaign is identified as spam, is because
attackers deceive in believing that there exists some problem
with their PC and charge money in return. Previous work has focused on the methods used by attackers to convince the victim and to make money~\cite{miramirkhani2017dial}. In this paper, we are interested in looking at the cross-platform behavior of such tech support scam
campaigns.

Over the course of six months of data collection, we got a total of 43,552 posts spread across all the five OSNs propagating to the extent of 41 phone numbers. 
The complete dataset description for tech support campaigns is shown in Table~\ref{techsupport}. 
\begin{table}[h]
\centering
\caption{Statistics for Tech Support Campaign}
\label{techsupport}
\footnotesize
\setlength\tabcolsep{3pt}
\begin{tabular}{| l | P{1.16cm} | P{0.66cm} | P{0.66cm} | P{0.65cm} | P{0.66cm} |} \hline
Features & TW & FB & G+ & YT & FL \\ \hline\hline
Total Posts & 28,984 & 2,151 & 7,830 & 2,850 & 1,737\\ \hline
Posts with URLs & 25,245 & 1,391 & 5,714 & 227 &1,503\\ \hline
Distinct Phone Numbers & 41 & 33 & 37 & 39 & 20\\\hline
Distinct User IDs & 748 & 289 & 360 & 433 & 79\\\hline
Distinct Posts & 16,142 & 1,797 & 6,570 & 2,050 & 1,449\\\hline
Distinct URLs & 68 & 951 & 3,189 & 80 & 293\\\hline
\end{tabular}
\end{table}

As phone numbers are one of the primary tokens used by spammers, we examined carrier information tied to each number to identify what kind of phone numbers spammers use viz. landline, mobile, VoIP, or toll-free).
We derived this information from several online services like Twilio (mobile carrier information)~\footnote{https://www.twilio.com/}, Truecaller (spam score assigned to the phone number)~\footnote{http://truecaller.com/}, and HLR lookups (current active location of the phone number).~\footnote{https://www.hlr-lookups.com/} We found that all the phone numbers used in the Tech Support campaign were toll-free numbers. 
Using a toll-free number offers several advantages to a spammer: 
(1) increased credibility: it does not incur a cost to the person calling, hence people perceive it to be legitimate, 
(2) it provides international presence: spammers can be reached from any part of the world.
Further, we found that spammers used services like ATL, Bandwidth, and, Wiltel Communications to obtain these toll-free numbers and that a majority of them were registered between 2014 and 2016.

\subsection{How does Content Cross-pollinate?}
Now, we answer the following question: \textit{Is a particular OSN preferred to start the spread of a campaign?} \textit{Is there a specific pattern in the way spam propagates on different OSNs?}

\begin{figure}[h]
\subcapraggedrighttrue
\begin{center}
\subfigure[Posts across OSNs]
{ \label{fig:temporal_tech}
\includegraphics[width=0.45\linewidth]{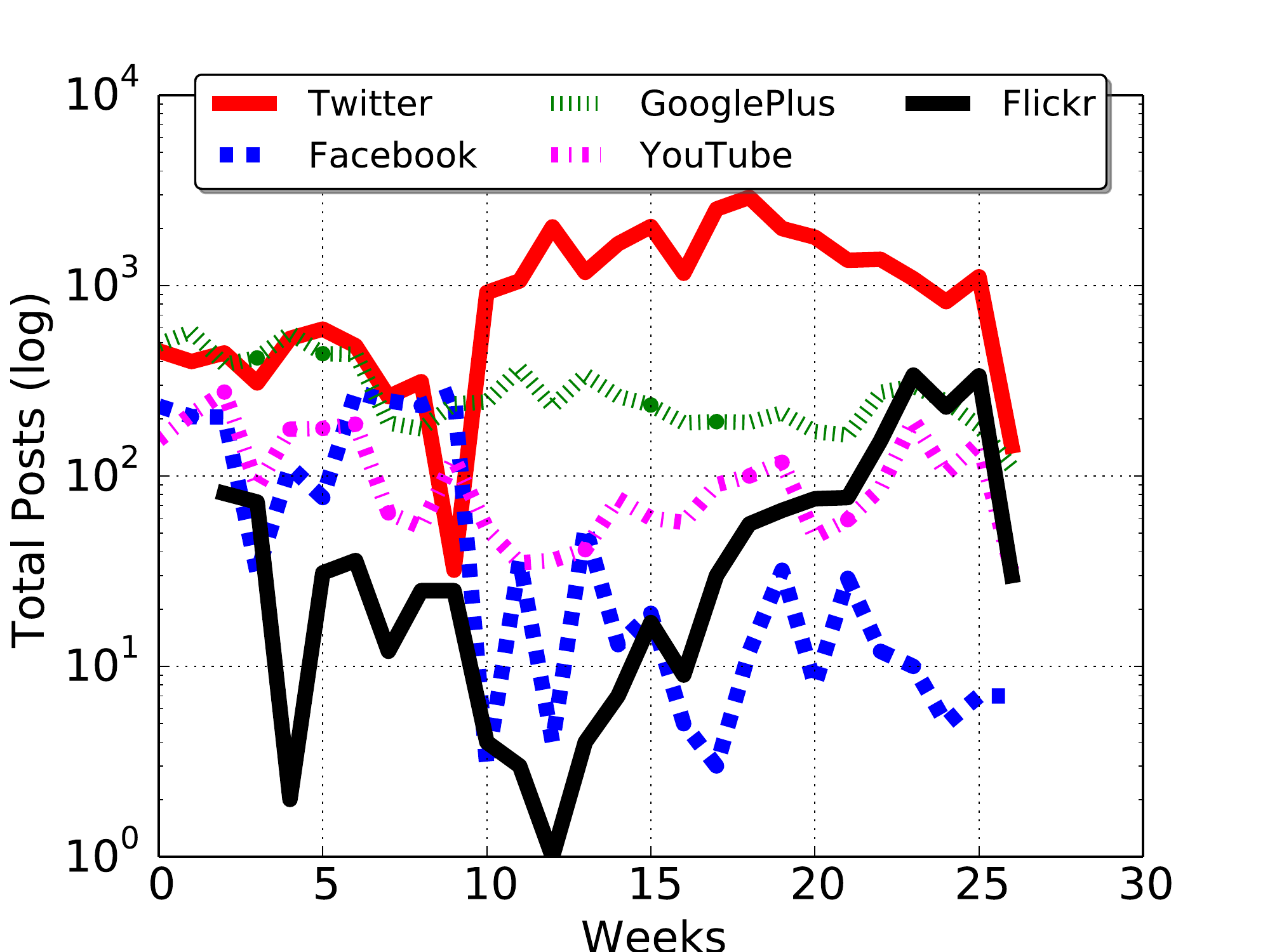}
}
\subfigure[Inter-arrival Time of Posts appearing on OSNs]
{ \label{fig:interarrival_tech}
\includegraphics[width=0.45\linewidth]{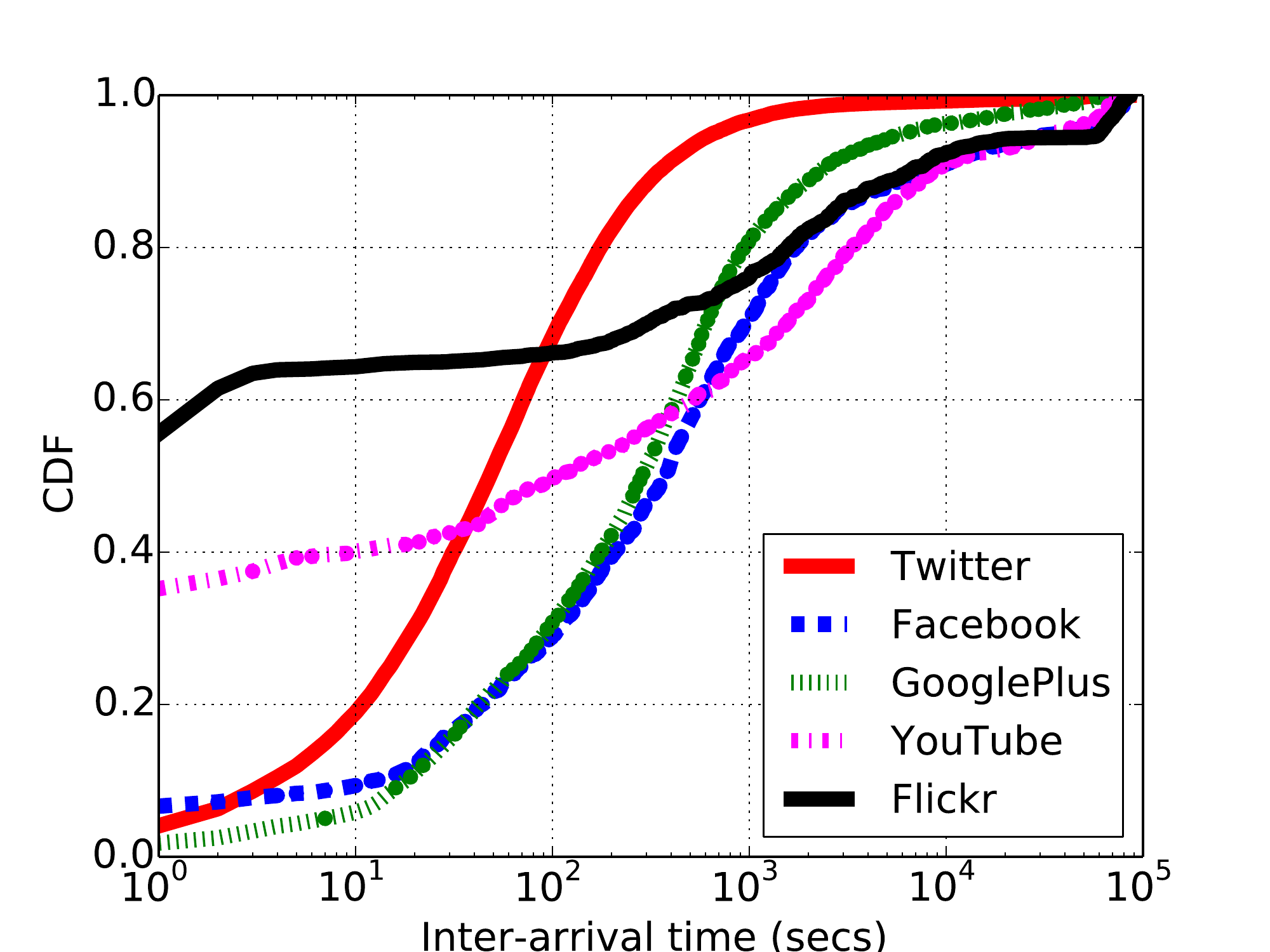}
}
\caption{Temporal properties of Tech Support Campaign across OSNs -- all OSNs are abused to spread the campaign but volume is maximum on Twitter. Inter-arrival time between two consecutive posts is minimum for Twitter. Spammers began to heavily abuse Flickr towards the end of our data collection.}
\label{fig:allcampaigns}
\end{center}
\end{figure}

Figure~\ref{fig:temporal_tech} shows the temporal pattern of content across OSNs. Note that our data collection was done over a period of six months while a campaign may have existed before and~/~or after this period. Hence, while the longest detected active time for a campaign in our dataset is 186 days, the actual time may be greater.
A majority of these posts are densely packed into a small number of short time bursts, while the entire campaign spans a much longer period.
Though the volume of content is significantly higher on Twitter, all OSNs are consistently being abused for propagation. Inter-arrival time, i.e., the average time between two successive posts is observed to be least on Twitter (308s), as shown in Figure~\ref{fig:interarrival_tech}. 
It is interesting to note that a few campaigns on Flickr have an inter-arrival time between two posts close to 1s, even though the average inter-arrival time is highest on Flickr. As Figure~\ref{fig:temporal_tech} shows, the volume on Flickr increased during the last few weeks of our data collection period. We divided the inter-arrival time into two time windows; first 15 weeks, and last 11 weeks. We observed that the average inter-arrival time in latter time window dropped from 9786s to 2543s which means spammers had started heavily abusing Flickr to spread the Tech Support campaign.
It is hard to ascertain the motivation of the spammers in sending high volume content on Twitter, but, we speculate one of the reasons could be the public nature of the Twitter platform, as compared to closed OSNs like Facebook. 
For all the phone numbers, we analyzed the appearance of phone numbers on different OSNs, and the order in which they appear, as reported in Table~\ref{tech_sequence}. 
\begin{table}[h]
\footnotesize
\centering
\caption{Distribution of phone numbers according to their first appearance amongst OSNs. Flickr is never chosen as a starting point and there is no particular sequence in which spam propagates across OSNs.}
\label{tech_sequence}
\begin{tabular}{|l|c|l|}
\hline
Starting OSN & \#Cases & Most common sequence \\ \hline \hline
Twitter (TW)      & 12      & TW $\rightarrow$ G+ $\rightarrow$ YT                                 \\ \hline
GooglePlus (G+)   & 10      & G+ $\rightarrow$ TW $\rightarrow$ YT $\rightarrow$ FB $\rightarrow$ FL     \\ \hline
Facebook (FB)    & 6       & FB $\rightarrow$ G+ $\rightarrow$ TW $\rightarrow$ YT                 \\ \hline
YouTube (YT)     & 13      & YT $\rightarrow$ G+ $\rightarrow$ TW $\rightarrow$ FB                      \\ \hline
\end{tabular}
\end{table}
For each network that is picked as the starting point, we identified the most common sequence in which phone numbers appeared subsequently on other OSNs. We found that Flickr was \emph{never} chosen as the starting OSN to initiate the spread of a phone number.
Further, we noticed that the posts originating from YouTube took the maximum time to reach a different OSN with an average inter-OSN time of 5 hours.

To summarize, we observed that all OSNs were abused to spread the Tech Support campaign, and no particular OSN was preferred to drive the campaign. In addition, there was no particular sequence in which spam propagated across OSNs.

\subsection{How do Spammers Maximize Visibility?}
We observed various strategies adopted by spammers to increase the dissemination of their posts. In this section, we discuss those strategies and their effectiveness. 

The \emph{Visibility} of a post is defined as the action performed by the user (consumer of the post) in terms of liking or sharing the post, which accounts for traction a particular post received. 
For each network, we define the value of visibility as follows:  number of likes and reshares on Facebook, +1s and reshares on GooglePlus, number of likes and retweets on Twitter, and video like count on YouTube. 
We did not consider Flickr in our analysis since Flickr API gives only the view count of the image posted on the platform. A user only viewing an image cannot be assumed to be a victim of the campaign. 
To calculate visibility in all scenarios, we collected the \emph{likes~/~retweets, plus-oners~/~reshares,} and \emph{likes} from Twitter, GooglePlus, and Facebook respectively using their APIs. Apart from calculating values for each visibility attribute, we also collected properties of the user accounts involved, i.e., the IDs of user accounts involved in retweeting~/~liking~/~resharing the content. Due to rate limiting constraints on  each of the APIs, we could not fetch visibility information daily. We collected this data six months after our data collection period, as posts take time to reach their audience. Due to this, (1) we might have missed information of tweets posted by suspended accounts, and (2) our total visibility values represent a lower bound.

To increase the visibility of content, we observed that the spammers use the following tricks: 67\% of posts contained hashtags  (for marketing~\cite{carrascosa2013trending}, gaining followers~\cite{martin2016hashtags}),
82.7\% of posts contained URLs (for increased engagement with potential victims), 
12.1\% of posts contained short URLs (for obfuscating the destination of a URL and getting user engagement analytics), and 72\% of posts contained photos (as visual content gathers more attention). 
We also noticed collusion between accounts and cross-referenced posts to increase the visibility of the campaign.

\textbf{Cross-referenced posts:} We call a post cross-referenced if it was posted to OSN X, but contains a URL redirecting to OSN Y.
For instance, a Twitter post containing a link `fb.me/xxxx' which would redirect to a different OSN, Facebook. 
Spammers either direct victims to existing posts or to another profile which is propagating the same campaign on a different OSN. 
In the Tech Support campaign, we observed that 
3.2\% of Facebook posts redirected to YouTube, and 1.78\% of posts redirected from GooglePlus to YouTube. 

\textbf{Collusion between accounts:} 
In the Tech Support campaign, we observed traces of collusion, i.e., spammers involved in a particular campaign, \emph{like / share} each other's posts on OSNs or like their content to increase reachability.
Collusion helps in cascading information to other followers in the network.

We calculated the visibility received by all the posts after removing likes~/~reshares~/~retweets by the colluders (i.e., accounts spreading the campaign already present in the dataset). 
We noticed that the posts containing the above-mentioned attributes (hashtags, URLs, short URLs, photos, cross-referencing, and collusion) garnered around ten times more visibility than posts not containing them.
Around 10\% of the posts saw traces of collusion, contributing to 20\% of the total visibility. Maximum visibility (22.1\% of total visibility) was observed  for posts containing hashtags. 
In addition, we observed that a major chunk of visibility came from GooglePlus, followed by Facebook. 
This shows that the audience targeted influences the visibility garnered by a particular campaign, as GooglePlus is known to be consumed mostly by IT professionals~\footnote{\url{https://insight.globalwebindex.net/chart-of-the-day-who-is-most-likely-to-use-google}}.

\subsection{To what Extent OSNs Suspend User Accounts?}\label{useridspool}
\begin{figure}[h]
\subcapraggedrighttrue
\begin{center}
\subfigure[New users created from time to time for campaign sustainability.]
{ \label{fig:newusers_tech}
\includegraphics[width=0.45\linewidth]{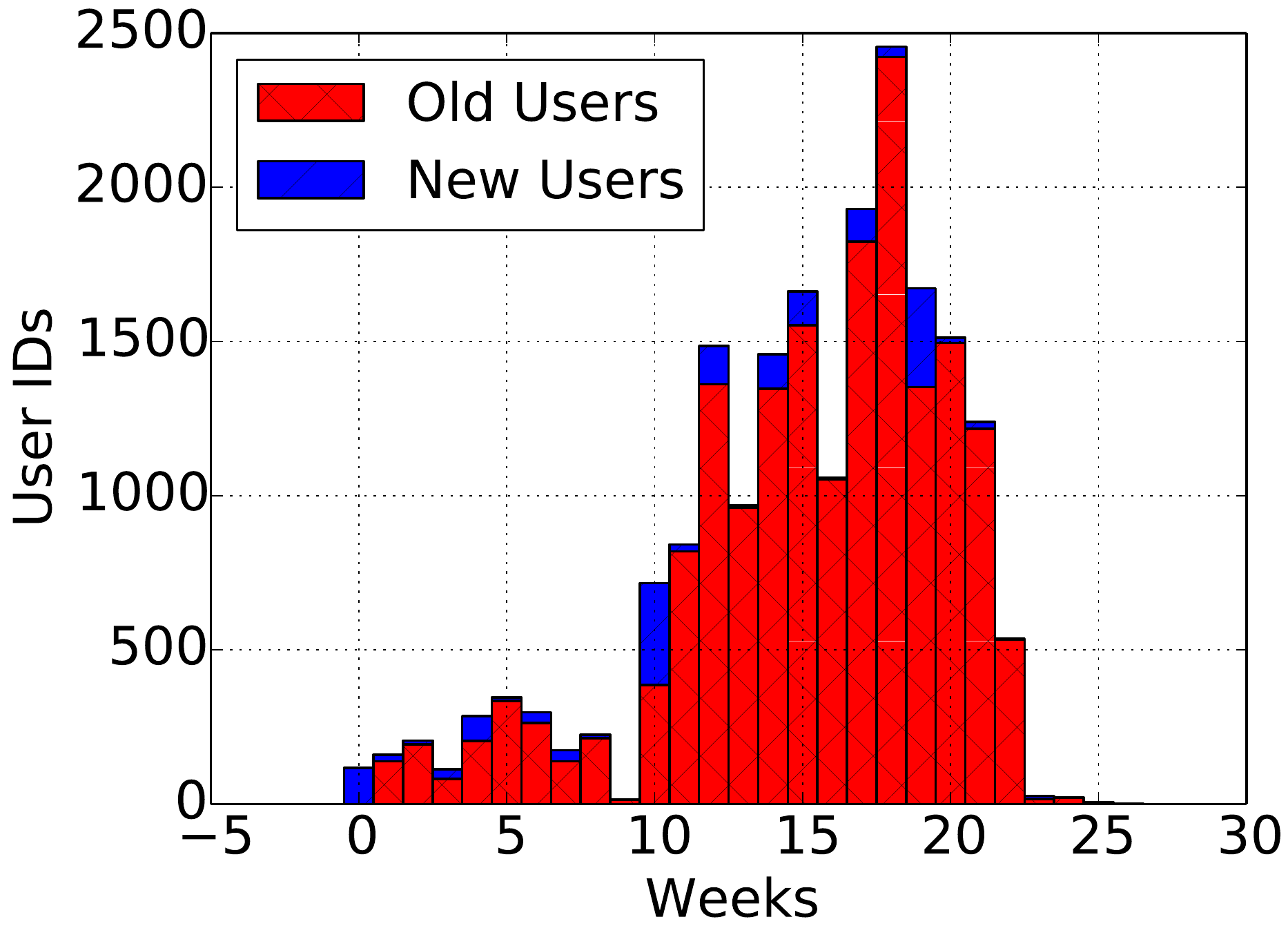}
}
\subfigure[Volume per user kept low to evade suspension.]
{ \label{fig:vol_user_tech}
\includegraphics[width=0.45\linewidth]{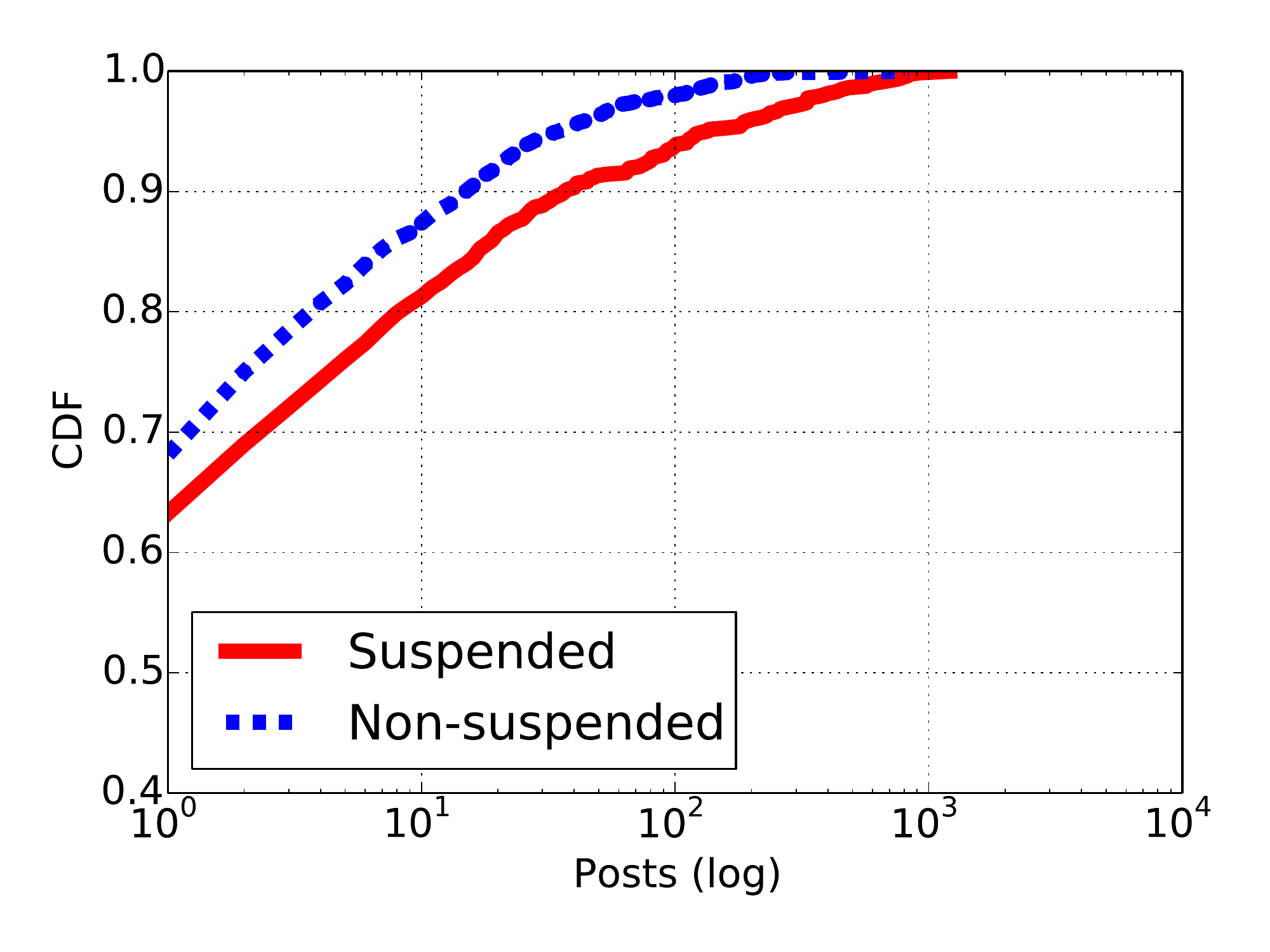}
}
\caption{New user accounts created from time to time and volume per ID kept low, to avoid suspension in the Tech Support Campaign.}
\label{fig:allcampaigns}
\end{center}
\end{figure}

To aid in the propagation of a campaign, spammers
manage multiple accounts to, garner a wider audience,
withstand account suspension, and in general increase the volume. Individual spammer accounts can either use automated techniques to aggressively post about a campaign or use hand-crafted messages.
In this section, we examine the behavior of user accounts behind the Tech Support campaign.
Spammers want to operate accounts in a stealth mode, which requires individual accounts to post few posts. It costs effort to get followers to a spam account, and the number of `influential' accounts owned by a spammer is limited. Thus, the spammer tends to repeatedly use accounts to post content keeping volume low per account (Figure~\ref{fig:vol_user_tech}), while creating new accounts once in a while (Figure~\ref{fig:newusers_tech}).

\textbf{Long-lived user accounts:}
During our data collection, we found that 68.7\% (1,305) of the accounts were \emph{never} 
suspended or taken down on any of the five OSNs.
This is in stark contrast to the URL based campaigns~\cite{thomas2011suspended},
where the authors observed that 92\% of the user accounts were suspended within three days of their first tweet.
To take into account delays in the OSNs' account suspension algorithm, 
we queried all the accounts six months after the data collection to determine which accounts were deleted~/~suspended. This process consists
of a bulk query to each OSN's API with the profile ID of the account.~\footnote{If the account is deleted~/~suspended, (a)
Twitter redirects to \url{http://twitter.com/suspended}, and returns error 404, (b) 
Youtube returns `user not found', (c) Facebook returns error 403 in case the account is suspended, (d)  GooglePlus throws a 
`not found' error, (e) Flickr responds with a `user not found' error.} 
For each of these accounts, we looked at the time stamp of the first and last post within our dataset, after which we assumed that the account was suspended immediately. 
Out of the accounts which were suspended, around 35\% of the accounts were suspended within a day of their first post; the longest lasting account was active
for 158 days, before finally getting suspended. 
On an average, accounts got suspended after being active for 33 days. This is in clear contrast to users getting suspended within three days for URL based spam campaigns, and thus, focused efforts are needed to strengthen defense from evolving phone-based spam campaigns.

\subsection{Is Existing Intelligence based on URLs Useful to Handle Phone-based Spam?}\label{urlsdomains}
Apart from creating accounts to propagate content, and using phone numbers to interact with victims, spammers also need a distinct set of URLs to advertise.
In this section, we look at the domains, subdomains and URL shorteners used by spammers. Of all the posts, we had 4,581 unique URLs and 594 distinct domains. Of all the URLs, 12.1\% were shortened using bit.ly; 3\% of them received over 69,917 clicks (data collected from bit.ly API), showing that the campaign was fairly successful. 

Given the prevalence of spam on OSNs, we examined the effectiveness of existing blacklists to detect malicious domains. Specifically, we used Google safe browsing~\footnote{https://developers.google.com/safe-browsing/v4/lookup-api} 
and Web of Trust (WOT)~\footnote{https://www.myWOT.com/wiki/API} to see if they were effective in flagging domains as malicious. 
Web of Trust categorizes the domains into several reputation buckets along with the confidence to assign a category. Please note that one domain may be listed in multiple categories.
We marked a domain as malicious if the domain appeared in any of the following categories -- 
negative (malware, phishing, scam, potentially illegal), questionable (adult content). We checked the URLs and domains even after six months of data collection since blacklists may be slow in updating response to new spam sites. 
We marked a URL malicious if it was listed as malicious either by Google safe browsing or WOT. We checked these domains against the blacklists, finding that 10\% of the domains were blacklisted by WOT, 
none by Google safe browsing. Overall, we found that existing URL infrastructure was ineffective to blacklist URLs used in phone-based spam campaigns.


\subsection{Can Cross-Platform Intelligence be used?}
Given that existing URL infrastructure is ineffective, we study if cross-platform intelligence across OSNs can be used. To this end, we look at the spam user profiles across OSNs to figure out which OSN is most effective in building the intelligence.

\textbf{Homogeneous identity across OSNs:}
Simply analyzing users' previous posts might not be sufficient, as users can switch between multiple identities, making it hard for OSN service providers to detect and block them. Moreover, spammers may appear legitimate based on the small number of posts made by a single identity. 
The challenge remains in analyzing the aggregate behavior of multiple identities. 
To understand how user activity is correlated across OSNs, we pose the question: \textit{do users have a unique identity on a particular OSN or do they share identities across OSNs? Within the same network, can we find the same users sharing multiple identities?} 

To answer this, we looked at user identities across different OSNs in \emph{aggregate} 
(multiple identities of the same user across different OSNs) and 
\emph{individual} (multiple identities of the same user on a single OSN) forms.
If the \emph{same} user has multiple identities, sharing similar name or username, it is said to exhibit a homogeneous identity. 
To define user identity in a particular campaign, we used two textual features: \emph{name} and \emph{username}~\cite{ottoni2014pins}. 
Since networks like YouTube and Google Plus do not provide the username, 
we restrict matching to identities sharing the same name. 
We used Levenshtein distance to find similarity in usernames. LD($s_i, s_j $) is the Levenshtein edit distance between usernames $s_i$ and $s_j$. 
Here, LD($s_i, s_j$) = 1 means the strings are identical, while LD($s_i, s_j$ ) = 0 means they are completely different.
After manual verification by comparing profile images across OSNs, we found users having LD $>=$ 0.7 are homogeneous identities. 
We found four cases where multiple user identities were found for the same user within the same network, 
and in 65 instances, multiple user identities were present for the same user in more than two networks. 
Specifically, we found 51 users sharing multiple identities across two different OSNs, and 10 users sharing multiple identities across 3 OSNs.
We noticed that these accounts shared same phone numbers across OSNs; some accounts post more phone numbers that are part of tech support campaign. 

We found that the total number of posts made by these accounts was highest on GooglePlus (2696), followed by Twitter (1776), Facebook (577), Flickr (387), and YouTube (323). 
Out of all the homogeneous identities, the following are the percentages of accounts suspended on each OSN -- Twitter (60\%), YouTube (48\%), GooglePlus (32\%) Flickr (33\%), and Facebook (4\%). 
Our data is insufficient to determine whether account suspension is due to dissemination of content across OSNs or other unobserved spammers' properties. 
Notwithstanding, the association between user identities across OSNs, strengthens the fact that sharing information about spammer accounts across OSNs could help OSNs to detect spammers accurately.

\textbf{Reducing financial loss and victimization}: The actual number of users that are impacted depends on how many victims called spammers and bought the products advertised by campaigns. Since it is hard to get this data, we provide a rough estimate of the number of victims falling for campaigns identified in our dataset.
We find reputation of spammers in terms of their followers count on Twitter, friends~/~page likes on Facebook, circle count on GooglePlus, and subscriber count on Youtube. As these users have subscribed to spammers to get more content, they are likely to fall for the spam. Some of the users would be the ones who aren't aware of the campaign being spam, while some followers~/~friends could be spammers themselves who have followed other spammers' accounts. We again collected this data after 6 months of our data collection and recorded 637,573 followers on Twitter, 21,053 friends on Facebook, 11,538 followers on GooglePlus, and 2,816 likes on YouTube amounting to a total of 670,164 users. 
Please note that this number is a lower bound, as we were not able to retrieve statistics for suspended~/~deleted accounts.
Assume that we transfer knowledge from Twitter to other OSNs and prevent the onset of campaigns on other OSNs, we analyzed how much money and victims could be saved. 
Looking only at the friends, followers, and likers on Facebook, GooglePlus, and YouTube respectively, we could save 35,407 (21,053 + 11,538 + 2,816) unique victims and \$8.8M (35,407 * \$290.9) by transferring intelligence across OSNs. We used the average cost of the Tech Support Spam to be \$290.9 per victim, as reported by Miramirkhani et al.~\cite{miramirkhani2017dial}.
\section{Discussion} \label{discussion}
In this section, we provide a synthesis of our evaluations and propose some recommendations to OSN service providers.

\paragraph{\textbf{How spammers can be choked?}} Phone numbers are a stable resource for spam since spammers need to provide their real phone numbers so that victims can reach out to them. A solution built around phone numbers, therefore, would be more reliable in bringing down spammers. 
As a countermeasure, there are two potential mechanisms -- a) phone blacklist and b) suspension of OSN accounts. A~\emph{phone blacklist} should be created, similar to URL blacklists, to check if a phone number is involved in a spam~/~scam campaign. Blacklisting a phone number would break the connecting link between victims and spammers, thus bringing down the spammers' monetization infrastructure. However, it is difficult to create one, because there are little identifiable features associated with a phone number as there are with URLs like landing page, some special characters, domain typo-squatting, etc.
Therefore, user suspension which can be collected from OSNs can come to rescue. From this research we established that the link between a phone number and the spammer account is crucial. Thus, one can focus on removing malicious users from user communities sharing the same phone number. In this network of user accounts, some users would already be suspended by OSNs. The labels can be recursively propagated to other unknown nodes from the known suspended nodes using several graph-based algorithms like Page Rank. Bringing down the spammers propagating phone numbers would disintegrate the entire campaign. 

There exist some services, like Truecaller~\footnote{\url{https://www.truecaller.com/}} and FTC's do-not-call complaint dataset~\footnote{\url{https://www.ftc.gov/site-information/open-government/data-sets/do-not-call-data}}, which collect information about phone numbers that spammers use to call victims~(\textit{incoming spam communication}). In this work, however, we demonstrated that spammers advertise their phone numbers across OSNs, so that victims would call them instead~(\textit{outgoing spam communication}). We found the overlap between our collected phone numbers (associated with potential spam campaigns) with the FTC (0.001\%) and Truecaller (0.4\%) databases to be minimal. It is, therefore, imperative that solutions also be built on outgoing spam communication.

\paragraph{\textbf{Measuring Impact using Honeypots}}
In this work, we focused on using friends and followers of the user as a metric to measure the impact; it might not capture the actual victims who fell for those campaigns. 
As an alternative approach, one can simulate a campaign; changing the phone number (say to phone number X) and keeping the text intact. There are certain services like Twilio~\footnote{https://www.twilio.com/} that aid in making calls over the Internet, which can be used to record the number of calls being made to phone number X. Spammer networks are dense; to ensure that these simulated campaigns are visible to a large OSN population, one can use Facebook Ads~\footnote{https://www.facebook.com/about/ads} or Twitter Ads~\footnote{https://business.twitter.com/en/twitter-ads.html} for campaign promotion as advertisements.
We believe this is a potential way to measure the impact of campaigns. 
\section{Conclusion} \label{conclusion}
With the convergence of telephony and the Internet, the phone channel has become an attractive target for spammers to exploit and monetize spam conducted over the Internet.
This paper presents the first large-scale study of cross-platform spam campaigns that abuse phone numbers. We collect $\sim$23 million posts containing $\sim$1.8 million unique phone numbers from Twitter, Facebook, GooglePlus, Youtube, and Flickr over a period of six months. We identified 202 campaigns running from all over the world with Indonesia, United States, India, and the United Arab Emirates being the highest contributors. We showed that even though Indonesian campaigns generated $\sim$3.2 million posts, only 1.6\% have been suspended so far. However, the number of accounts suspended in a campaign is not correlated with volume. Campaigns providing escort services and technical support solutions had more account suspensions. After interacting with spammers, we observed that they adopt tactics similar to legitimate services, to convince victims. By examining campaigns running across OSNs, we showed that Twitter could suspend $\sim$93\% more accounts spreading spam as compared to Facebook. Therefore, sharing intelligence about spam user accounts across OSNs can aid in spam detection; $\sim$35K victims and \$8.8M could be saved based on exploratory analysis of our data. 
We acknowledge that our validations on some possible explanations proposed in this work may be not rigorous, due to difficulties in thoroughly obtaining spammers' motivations. However, we believe that our first-of-its-kind analysis of these phenomena still provides great value and opens new doors to understand the phone-based spammer ecosystem across OSNs better.

\bibliographystyle{ACM-Reference-Format}
\bibliography{ref} 
\section{Appendix}
\subsection{Regular Expressions for Data Collection}\label{regex}
We used a curated list of 400 keywords like call, SMS, WhatsApp, ring, contact, dial, reach etc to filter relevant tweets from Twitter's Streaming API. While extracting phone numbers from the tweets, we encountered variations in representation of phone numbers, for instance the number 1-888-551-2881 can be represented
as 1(888)551-2881, 1(888) 551-2881, 1.888.551.2881,
or 1 888 551 2881 where all variations were being counted as different phone numbers. We filtered out this noise by post-processing the data, where a couple of regular expressions were used to obtain a valid phone number from the text obtained from each post, some of which are listed below. \\

\small
\begin{Verbatim}[frame=single]
1. ('(?<= )\d{6}-\d{3}(?= )|
(?<=\[)\d{6}-\d{3}(?=\])|(?<=\()\d{6}-\d{3}(?=\))')
2. (' (\d[\d ]{5,13}\d{2}) ')
3. ('\$ *\d+[\.]*\d+|\d+[\.]*\d+\$')
4. (ur'\xe2\x80\xa6')
5. ('^\d+\s|\s\d+\s|\s\d+$')
\end{Verbatim}
\normalsize

\subsection{Sample of Transcribed Calls with Spammers}\label{transcript}
\small
IVR: Press 1 to know about our products, 2 to check the status of previous order and 3 for other inquiries \\
Victim:  *pressed 1* \\
IVR: Press 1 to know more about <company-name> viagra pills and 2 for other products. \\
Victim: *pressed 1* \\
IVR: *call forwarding to human* \\
Scammer: Hello, I'm <name>, speaking from <company-name>, what would you like to know about the <brand> viagra pills. \\
Victim: What are the various packs I can buy and how much does it cost? \\
Scammer: We have only one variant which costs \textdollar99 - \textdollar119 for the pills and \textdollar20 for delivery.\\
Victim: Okay. How can I pay for the order if I decide to order? Do you have a web portal where I can make an online transaction? \\
Scammer: No sir, currently, we're operating only over phone, so you can provide your VISA card details to me, and I'll be happy to place the order for you. \\
Victim: Is phone the only option? I would like to make the payment through the web portal. \\
Scammer: Sorry sir, but we operate only over phone. \\
Victim: Okay, what are the product guarantees you offer? \\
Scammer: Yes, sir please be assured that we provide 100\% return guarantee. \\
Victim: Can I get some samples before placing the order? \\
Scammer: I am sorry sir, we don't provide samples. Should I place an order for you? \\
Victim - No, thank you for the information.

\end{document}